\documentclass[aps,floats]{revtex4}
\usepackage{amsmath,amssymb}
\usepackage{graphicx,epsfig}

\begin{document}
\bibliographystyle {plain}

\def\oppropto{\mathop{\propto}} 
\def\opsimeq{\mathop{\simeq}}
\def\opoverderline{\mathop{\overline}}
\def\operarrow{\mathop{\longrightarrow}}
\def\opsim{\mathop{\sim}}

\def\fig#1#2{\includegraphics[height=#1]{#2}}
\def\figx#1#2{\includegraphics[width=#1]{#2}}


\title{  Dyson Hierarchical Long-Ranged Quantum Spin-Glass \\
via real-space renormalization  } 


\author{ C\'ecile Monthus }
 \affiliation{Institut de Physique Th\'{e}orique, Universit\'e Paris Saclay, CNRS, CEA, 91191 Gif-sur-Yvette, France}

\begin{abstract}
We consider the Dyson hierarchical version of the quantum Spin-Glass with random Gaussian couplings characterized by the power-law decaying variance $\overline{J^2(r)} \propto r^{-2\sigma}$ and a uniform transverse field $h$. The ground state is studied via real-space renormalization to characterize the  spinglass-paramagnetic zero temperature quantum phase transition as a function of the control parameter $h$. In the spinglass phase $h<h_c$, the typical renormalized coupling grows with the length scale $L$ as the power-law $J_L^{typ}(h) \propto \Upsilon(h) L^{\theta}$ with the classical droplet exponent $\theta=1-\sigma$, where the stiffness modulus vanishes at criticality $\Upsilon(h) \propto (h_c-h)^{\mu} $, whereas the typical renormalized transverse field decays exponentially $ h^{typ}_L(h) \propto e^{- \frac{L}{\xi}}$ in terms of the diverging correlation length $\xi \propto (h_c-h)^{-\nu}$. At the critical point $h=h_c$, the typical renormalized coupling $J_L^{typ}(h_c) $ and the typical renormalized transverse field $ h^{typ}_L(h_c)$ display the same power-law behavior $L^{-z}$ with a finite dynamical exponent $z$. The RG rules are applied numerically to chains containing $L=2^{12}=4096 $ spins in order to measure these critical exponents for various values of $\sigma$ in the region $1/2<\sigma<1$.

\end{abstract}

\maketitle

\section{ Introduction }

In the field of classical spin-glasses, the Long-Ranged case of Gaussian couplings
where the variance decays as a power-law of the distance $r$
\begin{eqnarray}
\overline{ J^2(r) } = \frac{1}{r^{2 \sigma}}
\label{jrsigma}
\end{eqnarray}
has been much studied recently \cite{kotliar,BMY,KY,KYgeom,KKLH,KKLJH,Katz,KYalmeida,Yalmeida,KDYalmeida,LRmoore,KHY,KH,mori,wittmann,us_overlaptyp,us_dynamic,us_chaos,c_rgsgzerotemp,c_rgsgdyn,wittmann_dyn,alain} for various reasons : from the experimental point of view, the RKKY interaction in real spin-glasses actually decays as a power-law of the distance; from the numerical point of view, the Long-Ranged interaction can be studied in one dimension as a function of $\sigma$ on much larger sizes than
the case of Short-Ranged interaction as a function of the dimension $d$ ; from the theoretical point of view, the Long-Ranged case is also easier to analyze via scaling arguments.
 In particular, within the droplet scaling theory \cite{heidelberg,Fis_Hus}, the droplet exponent 
$\theta$ governing the scaling 
of the renormalized random coupling $J_L$ with the length $L$
 \begin{eqnarray}
J_L \propto L^{\theta}
\label{deftheta}
\end{eqnarray}
is expected to be given by the exact simple formula \cite{Fis_Hus,BMY} 
 \begin{eqnarray}
\theta^{LR}(d,\sigma) = d-\sigma
\label{thetaLR}
\end{eqnarray}
in the region where it is bigger than its short-ranged value $ \theta^{LR}(d,\sigma)> \theta^{SR}(d)$,
whereas the short-ranged droplet exponent  $\theta^{SR}(d)$ is known only numerically as a function of $d$ apart from the simple case $ \theta^{SR}(d=1)=-1$.
Since the spin-glass phase is stable at low-temperature when the droplet exponent is positive
$\theta>0$,
and since the ground state energy is extensive for $2 \sigma>d $ (Eq. \ref{jrsigma}),
one obtains that the interesting region where the spin-glass phase exists up to a finite temperature $T_c$
 \begin{eqnarray}
\frac{d}{2} < \sigma < d
\label{regionsigma}
\end{eqnarray}
exists already in dimension $d=1$. 

Besides the thermal transition of the classical spin-glass,
it is also interesting to consider the effect of quantum fluctuations introduced by a uniform transverse-field $h$ at zero temperature \cite{dutta}
\begin{eqnarray}
H= - \sum_{i}  h \sigma^z_i - \sum_{(i,j)} J_{i,j} \sigma_i^x \sigma^x_j 
\label{hqsg}
\end{eqnarray}
where the $J_{i,j}$ are the random couplings with the variance given by Eq. \ref{jrsigma}.
For $h=0$, one recovers the classical case with the droplet exponent of Eq. \ref{thetaLR},
so that the region of interest is the same as in Eq. \ref{regionsigma}. For large $h$, the ground-state will be paramagnetic, so that one expects a quantum phase transition at zero temperature at some critical transverse field $h_c$.
For the case of short-ranged couplings, the quantum spin-glass transition
is governed by an Infinite Disorder Fixed Point obtained via Strong Disorder Renormalization,
either exactly in dimension $d=1$ \cite{fisher} or numerically in dimensions $d=2,3,4$  \cite{fisherreview,motrunich,lin,karevski,lin07,yu,kovacsstrip,kovacs2d,kovacs3d,kovacsentropy,kovacsreview} 
(Note that this Infinite Disorder Fixed Point can also be reproduced via real-space Block Renormalization \cite{nishi,c_pacheco2d}). As a consequence,  the relation with the power-law mean-field theory
based on the properties of the infinite ranged model \cite{BM_quantumSK,miller} 
and supposed to apply in $d>8$ \cite{sachdev} has remained unclear. However for the case of 
 Long-Ranged couplings, the activated scaling of the renormalized couplings found 
at Infinite Disorder Fixed Point is not possible anymore, since the decay cannot be less than the initial power-law decay (Eq \ref{jrsigma}).
As a consequence, the dynamical exponent $z$ is finite, and one may expect more conventional
power-law critical properties. The case of the Long-Ranged {\it random ferromagnetic } quantum
 Ising chain
has been studied recently both via Strong Disorder Renormalization
 \cite{strongLR,strongLRbis,epiLR} and via Block Renormalization \cite{c_dysontransverse}.
The aim of the present paper is to study the properties of the Long-Ranged quantum spin-glass chain
 via the block real-space renormalization 
used previously for the ferromagnetic case \cite{c_dysontransverse}.

The paper is organized as follows.
In section \ref{sec_rg}, we describe the real-space renormalization rules for
 the Dyson hierarchical version of the quantum Long-Ranged Spin-Glass model,
and solve them deep in the spinglass phase and deep in the paramagnetic phase.
In section \ref{sec_nume}, the RG rules are applied numerically in the critical region
 to measure the critical behaviors.
Our conclusions are summarized in section \ref{sec_conclusion}.

\section{ Real-space renormalization approach }

\label{sec_rg}

\subsection{ Dyson hierarchical Quantum Long-Ranged Spin-Glass model }

Since the studies concerning the Dyson hierarchical classical ferromagnetic Ising model
\cite{dyson,bleher,gallavotti,book,jona,baker,mcguire,Kim,Kim77,us_dysonferrodyn},
many long-ranged disordered models have been analyzed via their Dyson hierarchical analogs
including for instance random fields Ising models \cite{randomfield,us_aval,decelle} and Anderson localization models \cite{bovier,molchanov,krit,kuttruf,fyodorov,EBetOG,fyodorovbis,us_dysonloc}.
Here we start from the Dyson hierarchical classical spin-glass model
 \cite{franz,castel_etal,castel_parisi,castel,angelini,guerra,barbieri} and we simply add
transverse fields to introduce quantum fluctuations.
More precisely, the Dyson hierarchical Quantum Spin-Glass model for $N=2^n$ spins
is defined as a sum over the generations $k=0,1,..,n-1$
\begin{eqnarray}
H_{(1,2^n)} &&  = \sum_{k=0}^{n-1} H^{(k)}_{(1,2^n)} 
\label{recDyson}
\end{eqnarray}
The Hamiltonian of generation $k=0$ contains the 
transverse fields $h_i$ and the lowest order couplings $J^{(0)}_{2i-1,2i}$ 
\begin{eqnarray}
 H^{(k=0)}_{(1,2^n)} &&  = - \sum_{i=1}^{2^n} h_i \sigma_i^z
 - \sum_{i=1}^{2^{n-1}}  J^{(0)}_{2i-1,2i}   \sigma_{2i-1}^x \sigma_{2i}^x
\label{h0dyson}
\end{eqnarray}
The Hamiltonian of generation $k=1$ reads
\begin{eqnarray}
 H^{(k=1)}_{(1,2^n)} &&  =
 - \sum_{i=1}^{2^{n-2}}  
\left[ J^{(1)} _{4i-3,4i-1}   \sigma_{4i-3}^x \sigma_{4i-1}^x
+  J^{(1)} _{4i-3,4i}   \sigma_{4i-3}^x \sigma_{4i}^x
+ J^{(1)} _{4i-2,4i-1}   \sigma_{4i-2}^x \sigma_{4i-1}^x
+  J^{(1)} _{4i-2,4i}   \sigma_{4i-2}^x \sigma_{4i}^x
 \right]
\label{h1dyson}
\end{eqnarray}
and so on up to the last generation $k=n-1$ that couples 
all pairs of spins between the two halves of the system
\begin{eqnarray}
 H^{(n-1)}_{(1,2^n)} = - \sum_{i=1}^{2^{n-1}} \sum_{j=2^{n-1}+1}^{2^n} 
 J^{(n-1)}_{i,j}  \sigma_i^x   \sigma_j^x 
\label{hlastdyson}
\end{eqnarray}

At generation $k$, associated to the length scale $L_k=2^k$,
the couplings $J^{(k)}_{i,j} $ read
\begin{eqnarray}
J^{(k)}_{i,j}=\Delta_k \epsilon_{ij}
\label{jndysonsg}
\end{eqnarray}
where $ \epsilon_{ij}$ are independent random variables of zero mean drawn with
the Gaussian distribution
 \begin{eqnarray}
G(\epsilon)  =  \frac{1}{\sqrt{4 \pi} } e^{- \frac{\epsilon^2}{4}}
\label{gaussian}
\end{eqnarray}
The characteristic scale $\Delta_k$ is chosen to decay exponentially
with the number $k$ of generations,
in order to mimic the power-law decay of Eq. \ref{jrsigma}
 with respect to the length scale $L_k=2^{k}$
\begin{eqnarray}
\Delta_k = 2^{-k \sigma} =  \frac{1}{L_k^{\sigma}} 
\label{deltandysonsg}
\end{eqnarray}
Then one expects that many scaling properties will be the same.
In particular, in the absence of the transverse fields, the classical ground state is
characterized by the same droplet exponent of Eq. \ref{thetaLR} for $d=1$
(see more details in \cite{c_rgsgzerotemp,c_rgsgdyn})
 \begin{eqnarray}
\theta(\sigma) = 1-\sigma
\label{thetaLR1}
\end{eqnarray}

Here we will consider the case where all the transverse field $h_i$ in Eq. \ref{h0dyson} are 
 all equal initially
\begin{eqnarray}
h_i=h
\label{uniformh}
\end{eqnarray}
so that this value $h$
 represents the control parameter of the spinglass-paramagnetic transition.
Upon renormalization however, the renormalized transverse field will become random
variables as we now describe.

\subsection{ Renormalization procedure }

We use the same renormalization procedure as in our previous work concerning the ferromagnetic case \cite{c_dysontransverse}. The only technical difference in the derivation of the RG rules is that we have to take into account the random amplitude and the random sign of the couplings. But of course the physical properties obtained from the RG flows will be completely different as expected for a spin-glass model. 

The elementary renormalization step concerns the box two-spin Hamiltonian of generation $k=0$ of Eq. \ref{h0dyson}
\begin{eqnarray}
H_{(2i-1,2i)} &&  = - h_{2i-1} \sigma_{2i-1}^z- h_{2i} \sigma_{2i}^z
- J^{(0)}_{2i-1,2i}   \sigma_{2i-1}^x \sigma_{2i}^x
\label{h1box}
\end{eqnarray}
Within the symmetric sector
\begin{eqnarray}
H_{(2i-1,2i)} \vert ++> && = - (h_{2i-1}+h_{2i}) \vert ++ >-  J^{(0)}_{2i-1,2i} \vert -- >
\nonumber \\
H_{(2i-1,2i)} \vert --> && =-  J^{(0)}_{2i-1,2i} \vert ++ >  +(h_{2i-1}+h_{2i})  \vert -- >
\label{huvs}
\end{eqnarray}
we project out the highest eigenvalue $\lambda_{2i}^{S+} = + \sqrt{ (J^{(0)}_{2i-1,2i})^2+(h_{2i-1}+h_{2i})^2 } $ to keep only the lowest eigenvalue
\begin{eqnarray}
\lambda_{2i}^{S-} = - \sqrt{ (J^{(0)}_{2i-1,2i})^2+(h_{2i-1}+h_{2i})^2 }
\label{lambdas}
\end{eqnarray}
corresponding to the eigenvector
\begin{eqnarray}
\vert \lambda_{2i}^{S-} > && = \cos \theta_{2i}^S\vert ++>+\sin \theta_{2i}^S\vert -- >
\label{vlambdas}
\end{eqnarray}
in terms of the angle $\theta_S$ satisfying
\begin{eqnarray}
\cos ( \theta_{2i}^S ) && = 
\sqrt{ \frac{1+ \frac{h_{2i-1}+h_{2i}}{ \sqrt{ (J^{(0)}_{2i-1,2i} )^2+(h_{2i-1}+h_{2i})^2 }}}{2}}
\nonumber \\
\sin ( \theta_{2i}^S ) && = {\rm sgn} (J^{(0)}_{2i-1,2i})
\sqrt{ \frac{1- \frac{h_{2i-1}+h_{2i}}{ \sqrt{ (J^{(0)}_{2i-1,2i} )^2+(h_{2i-1}+h_{2i})^2 }}}{2}}
\label{thetas}
\end{eqnarray}
Similarly within the antisymmetric sector
\begin{eqnarray}
H_{(2i-1,2i)} \vert +-> && = - (h_{2i-1}-h_{2i}) \vert +- >-  J^{(0)}_{2i-1,2i} \vert -+ >
\nonumber \\
H_{(2i-1,2i)} \vert -+> && =-  J^{(0)}_{2i-1,2i} \vert +- >  +(h_{2i-1}-h_{2i})  \vert -+ >
\label{huva}
\end{eqnarray}
we project out the highest eigenvalue $\lambda_{2i}^{A+} = + \sqrt{ (J^{(0)}_{2i-1,2i} )^2+(h_{2i-1}-h_{2i})^2 }$  to keep only the lowest eigenvalue
\begin{eqnarray}
\lambda_{2i}^{A-} = - \sqrt{ (J^{(0)}_{2i-1,2i} )^2+(h_{2i-1}-h_{2i})^2 }
\label{lambdaa}
\end{eqnarray}
with the corresponding eigenvector 
\begin{eqnarray}
\vert \lambda_{2i}^{A-} > && = \cos \theta_{2i}^A\vert +->+\sin \theta_{2i}^A\vert -+ >
\label{vlambdaa}
\end{eqnarray}
in terms of the angle $\theta_A$ satisfying
\begin{eqnarray}
\cos (  \theta_{2i}^A ) && =
\sqrt{ \frac{1+ \frac{h_{2i-1}-h_{2i}}{ \sqrt{ (J^{(0)}_{2i-1,2i} )^2+(h_{2i-1}-h_{2i})^2 }}
 }{2}}
\nonumber \\
\sin (  \theta_{2i}^A ) && = {\rm sgn} (J^{(0)}_{2i-1,2i})
\sqrt{ \frac{1- \frac{h_{2i-1}-h_{2i}}{ \sqrt{ (J^{(0)}_{2i-1,2i} )^2+(h_{2i-1}-h_{2i})^2 }}
 }{2}}
\label{thetaa}
\end{eqnarray}
In summary, for each two-spin Hamiltonian $H_{2i-1,2i}$ of Eq. \ref{h1box},
we keep only the two lowest states and label them as the two states 
of some renormalized spin $\sigma_{R(2i)}$
\begin{eqnarray}
\vert \sigma^z_{R(2i)}=+>  && \equiv \vert \lambda_{2i}^{S-} > 
\nonumber \\ 
\vert \sigma^z_{R(2i)}=->  && \equiv \vert \lambda_{2i}^{A-} > 
\label{sigmaR2states}
\end{eqnarray}
It is convenient to introduce the corresponding spin operators
\begin{eqnarray}
\sigma^z_{R(2i)} && \equiv \vert \sigma^z_{R(2i)}=+ > < \vert \sigma^z_{R(2i)}=+ \vert  - 
\vert \sigma^z_{R(2i)}=- > < \sigma^z_{R(2i)}=- \vert
\nonumber \\ 
\sigma^x_{R(2i)} && \equiv \vert \sigma^z_{R(2i)}=+ > < \vert \sigma^z_{R(2i)}=- \vert  + 
\vert \sigma^z_{R(2i)}=- > < \sigma^z_{R(2i)}=+ \vert
\label{opsigmaR}
\end{eqnarray}
and the projector
\begin{eqnarray}
P_{2i}^- \equiv \vert \sigma^z_{R(2i)}=+ > < \vert \sigma^z_{R(2i)}=+ \vert  + 
\vert \sigma^z_{R(2i)}=- > < \sigma^z_{R(2i)}=- \vert
\label{proj}
\end{eqnarray}

\subsection{ Renormalization rule for the transverse fields $ h_{R(2i)} $}

The projection of the Hamiltonian of Eq. \ref{h1box} reads
\begin{eqnarray}
P_{2i}^- H_{(2i-1,2i)} P_{2i}^- && =
 \lambda_{2i}^{S-} \vert \lambda_{2i}^{S-} > <\lambda_{2i}^{S-} \vert
 + \lambda_{2i}^{A-} \vert \lambda_{2i}^{A-} > <\lambda_{2i}^{A-} \vert
\nonumber \\
&& =\left( \frac{ \lambda_{2i}^{S-}+\lambda_{2i}^{A-}  }{2} \right)  P_{2i}^- 
  + \left( \frac{ \lambda_{2i}^{S-}-\lambda_{2i}^{A-}  }{2} \right)  \sigma^z_{R(2i)}
\label{projH}
\end{eqnarray}
so that the renormalized transverse field defined as the coefficient of $(-\sigma^z_{R(2i)} )$
is given by
\begin{eqnarray}
 h_{R(2i)} &&  = \left( \frac{ \lambda_{2i}^{A-}-\lambda_{2i}^{S-}  }{2} \right)
= \frac{ 2 h_{2i-1} h_{2i} }
{\sqrt{ (J^{(0)}_{2i-1,2i})^2+(h_{2i-1}+h_{2i})^2 }
+ \sqrt{ (J^{(0)}_{2i-1,2i})^2+(h_{2i-1}-h_{2i})^2 }}
\label{rgh}
\end{eqnarray}

\subsection{ Renormalization rules for the couplings $J_{R(2i),R(2j)} $}

The projection of the $\sigma^x$ operators
\begin{eqnarray}
P_{2i}^- \sigma^x_{2i-1} P_{2i}^- && = c_{2i-1} \sigma^x_{R(2i)}
\nonumber \\
P_{2i}^- \sigma^x_{2i} P_{2i}^- && = c_{2i} \sigma^x_{R(2i)}
\label{newbasis}
\end{eqnarray}
involves the two coefficients 
\begin{eqnarray}
c_{2i-1} && = {\rm sgn} (J^{(0)}_{2i-1,2i} )  \sqrt{ \frac{1+\frac{(J^{(0)}_{2i-1,2i})^2-h_{2i-1}^2+h_{2i}^2 }
{\sqrt{ (J^{(0)}_{2i-1,2i})^2+(h_{2i-1}+h_{2i})^2 }
\sqrt{ (J^{(0)}_{2i-1,2i})^2+(h_{2i-1}-h_{2i})^2 }}}{2}  }
\nonumber \\
c_{2i} && =  \sqrt{ \frac{1+\frac{(J^{(0)}_{2i-1,2i})^2+h_{2i-1}^2-h_{2i}^2 }
{\sqrt{ (J^{(0)}_{2i-1,2i})^2+(h_{2i-1}+h_{2i})^2 }
\sqrt{ (J^{(0)}_{2i-1,2i})^2+(h_{2i-1}-h_{2i})^2 }}}{2}  }
\label{coefcrg}
\end{eqnarray}
The renormalized coupling between $\sigma_{R(2i)} $ and $\sigma_{R(2j)} $
 is then given by the following linear combination of the four initial couplings
of generation $k$ associated to the positions $(2i-1,2i)$ and $((2j-1,2j)$
\begin{eqnarray}
 J_{R(2i),R(2j)} && = J^{(k)}_{2i,2j} c_{2i} c_{2j}
+ J^{(k)}_{2i-1,2j} c_{2i-1} c_{2j}
+ J^{(k)}_{2i,2j-1} c_{2i} c_{2j-1}
+ J^{(k)}_{2i-1,2j-1} c_{2i-1} c_{2j-1}
\label{rgjbox}
\end{eqnarray}

\subsection{ Iteration of the elementary renormalization step }

In summary, the elementary renormalization step described above maps the
initial model containing $L=2^n$ spins $(\sigma_1,...,\sigma_{2^n})$ with their transverse fields $h_i$ and their couplings $J^{(k)}_{i,j}$ into a renormalized model containing $\frac{L}{2}=2^{n-1}$ spins $(\sigma_{R(2)},...,\sigma_{R(2^{n})})$ with their renormalized transverse fields $h_{R(2i)}$ given by Eq. \ref{rgh}
and their renormalized couplings $J_{R(2i),R(2j)}$  given by Eq. \ref{rgjbox}.
The iteration is now straightforward : the next renormalization step will produce a renormalized model containing $\frac{L}{4}=2^{n-2}$ spins $(\sigma_{R^2(4)},...,\sigma_{R^2(2^{n})})$ with their renormalized transverse fields $h_{R^2(4i)}$
and their renormalized couplings $J_{R^2(4i),R^2(4j)}$, etc.
The renormalization procedure ends after $n$ RG steps where the whole sample 
containing $L=2^n$ initial spins has been renormalized into a single renormalized spin $\sigma_{R^n(2^{n})} $. 

\subsection{ RG flows deep in the Spin-Glass phase }

Deep in the Spin-Glass phase, the transverse fields $h_i$ are negligible with 
respect to the couplings $J_{i,j}$.
As a consequence, the coefficients of Eq. \ref{coefcrg} reduce to
\begin{eqnarray}
c_{2i-1} && \simeq {\rm sgn} (J^{(0)}_{2i-1,2i} ) 
\nonumber \\
c_{2i} && \simeq 1
\label{coefcrgsg}
\end{eqnarray}
and the RG rule for the renormalized coupling of Eq. \ref{rgjbox} becomes
\begin{eqnarray}
 J_{R(2i),R(2j)} && = J_{2i,2j} 
+ J_{2i-1,2j} {\rm sgn} (J^{(0)}_{2i-1,2i} ) 
+ J_{2i,2j-1} {\rm sgn} (J^{(0)}_{2j-1,2j} ) 
+ J_{2i-1,2j-1} {\rm sgn} (J^{(0)}_{2i-1,2i} ) {\rm sgn} (J^{(0)}_{2j-1,2j} ) 
\label{rgjboxrg}
\end{eqnarray}
This rule coincides with the RG rule for the classical spin-glass at zero temperature studied in detail
in \cite{c_rgsgzerotemp,c_rgsgdyn}, and in particular, the variance evolves according to
\begin{eqnarray}
 \overline{  (J^R_{R(2i),R(2j)})^2 }  \simeq 4 \overline{J_{2i,2j}^2 }
\label{varrgjboxsg}
\end{eqnarray}
In terms of the length $L=2^n$ obtained after $n$ RG steps, the variance grows
as
\begin{eqnarray}
&& \overline{  (J^R_L)^2 }  \simeq 4^n L^{-2 \sigma} = L^{2-2 \sigma} \equiv L^{2 \theta}
\label{varrgjboxsginte}
\end{eqnarray}
with the droplet exponent of Eq. \ref{thetaLR1}
\begin{eqnarray}
\theta =1- \sigma
\label{theta}
\end{eqnarray}
in agreement with the exact simple formula of Eq. \ref{thetaLR} 
predicted via scaling arguments \cite{Fis_Hus,BMY} and numerically
measured via Monte-Carlo \cite{KY}.

The RG rule of Eq. \ref{rgh} for the transverse field simplifies into
\begin{eqnarray}
 h_{R(2i)} && \simeq \frac{  h_{2i-1} h_{2i} }
{ \vert J^{(0)}_{2i-1,2i} \vert }
\label{rghstrong}
\end{eqnarray}
and is thus analogous to the usual Strong Disorder RG rule concerning
a strong-bond-decimation \cite{fisher}.
The iteration of this rule in log-variables
\begin{eqnarray}
\ln h_{R(2i)} && \simeq \ln  h_{2i-1}+ \ln   h_{2i} -\ln \vert J^{(0)}_{2i-1,2i} \vert 
\label{rghstronglog}
\end{eqnarray}
yields that the disorder average decays linearly in the length $L=2^n$ obtained after $n$ RG steps
\begin{eqnarray}
\overline{\ln h_{L}} && \simeq - (cst) L
\label{rghstronglogav}
\end{eqnarray}
whereas the variance grows as in the Central Limit theorem
\begin{eqnarray}
\overline{ (\ln h_{L})^2} - (\overline{\ln h_{L}})^2 && \propto L^{\frac{1}{2}}
\label{rghstronglogvar}
\end{eqnarray}

\subsection{RG flows  deep in the Paramagnetic phase }

Deep in the paramagnetic phase, the couplings $J_{i,j}$ are negligible with 
respect to the transverse fields $h_i$. As a consequence, the renormalized transverse fields
of Eq. \ref{rgh}
remain all equal to their common initial value $h$ (Eq. \ref{rgh})
\begin{eqnarray}
 h_{R(2i)} && \simeq h
\label{rghpara}
\end{eqnarray}
whereas the coefficients of Eq. \ref{coefcrg} simplify into
\begin{eqnarray}
c_{2i-1} && \simeq {\rm sgn} (J^{(0)}_{2i-1,2i} )  \frac{1}{\sqrt 2} 
\nonumber \\
c_{2i} && \simeq \frac{1}{\sqrt 2} 
\label{coefcrgpara}
\end{eqnarray}
so that the RG rule for the renormalized coupling of Eq. \ref{rgjbox} becomes
\begin{eqnarray}
 J_{R(2i),R(2j)} && =  \frac{1}{ 2}  \left[ J_{2i,2j} 
+ J_{2i-1,2j} {\rm sgn} (J^{(0)}_{2i-1,2i} ) 
+ J_{2i,2j-1} {\rm sgn} (J^{(0)}_{2j-1,2j} ) 
+ J_{2i-1,2j-1} {\rm sgn} (J^{(0)}_{2i-1,2i} ) {\rm sgn} (J^{(0)}_{2j-1,2j} ) \right]
\label{rgjboxrgpara}
\end{eqnarray}
In particular, the variance remains stable
\begin{eqnarray}
 \overline{  (J^R_{R(2i),R(2j)})^2 }  \simeq  \overline{J_{2i,2j}^2 }
\label{varrgjboxpara}
\end{eqnarray}
i.e. the renormalized couplings keep the same decay as the original couplings
\begin{eqnarray}
&& \overline{  (J^R_L)^2 }  \simeq  L^{-2 \sigma} 
\label{varrgjboxparainte}
\end{eqnarray}

\section{ Numerical study of the critical properties }

\label{sec_nume}

\begin{figure}[htbp]
 \includegraphics[height=4cm]{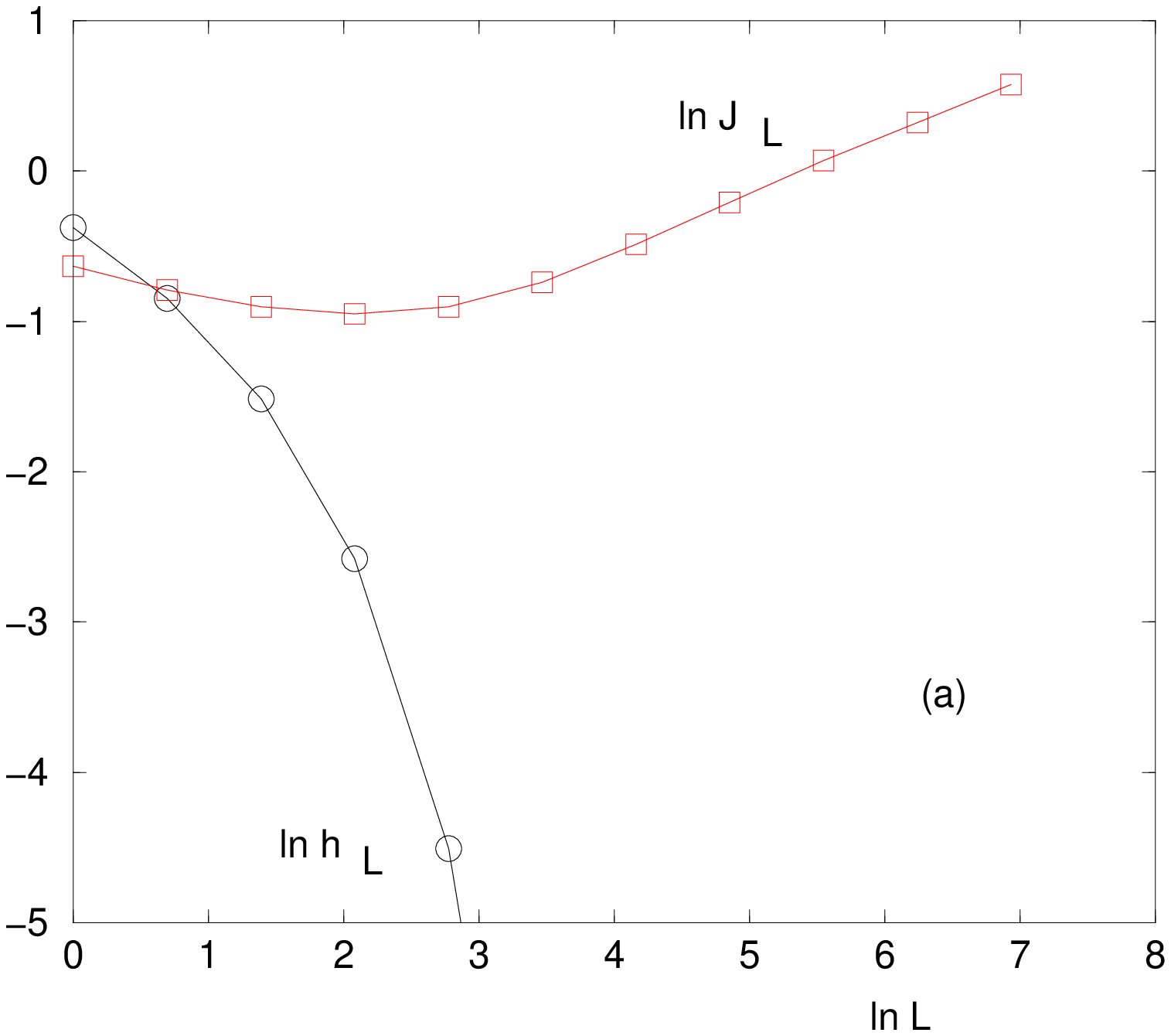}
\hspace{1cm}
 \includegraphics[height=4cm]{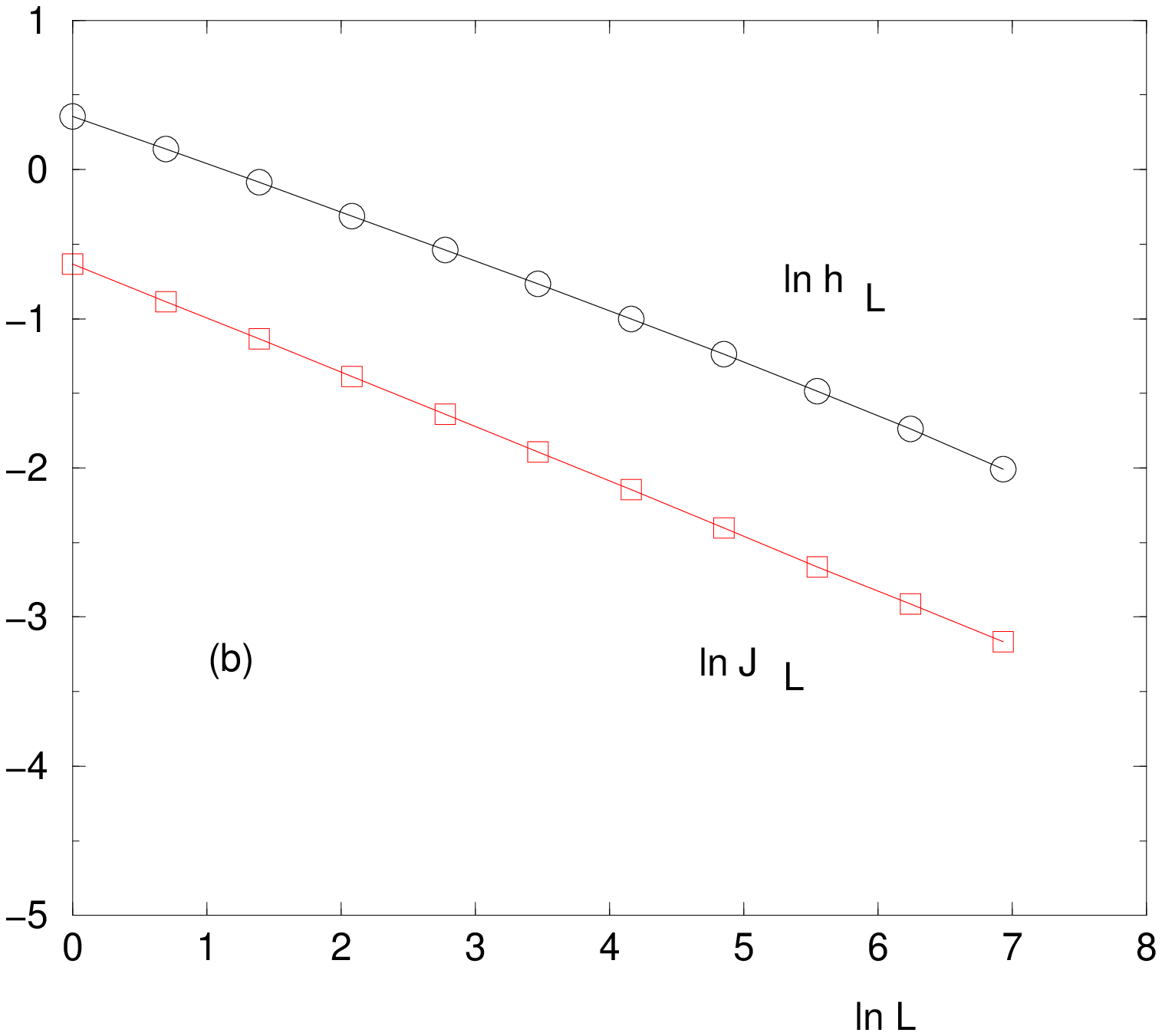}
\hspace{1cm}
 \includegraphics[height=4cm]{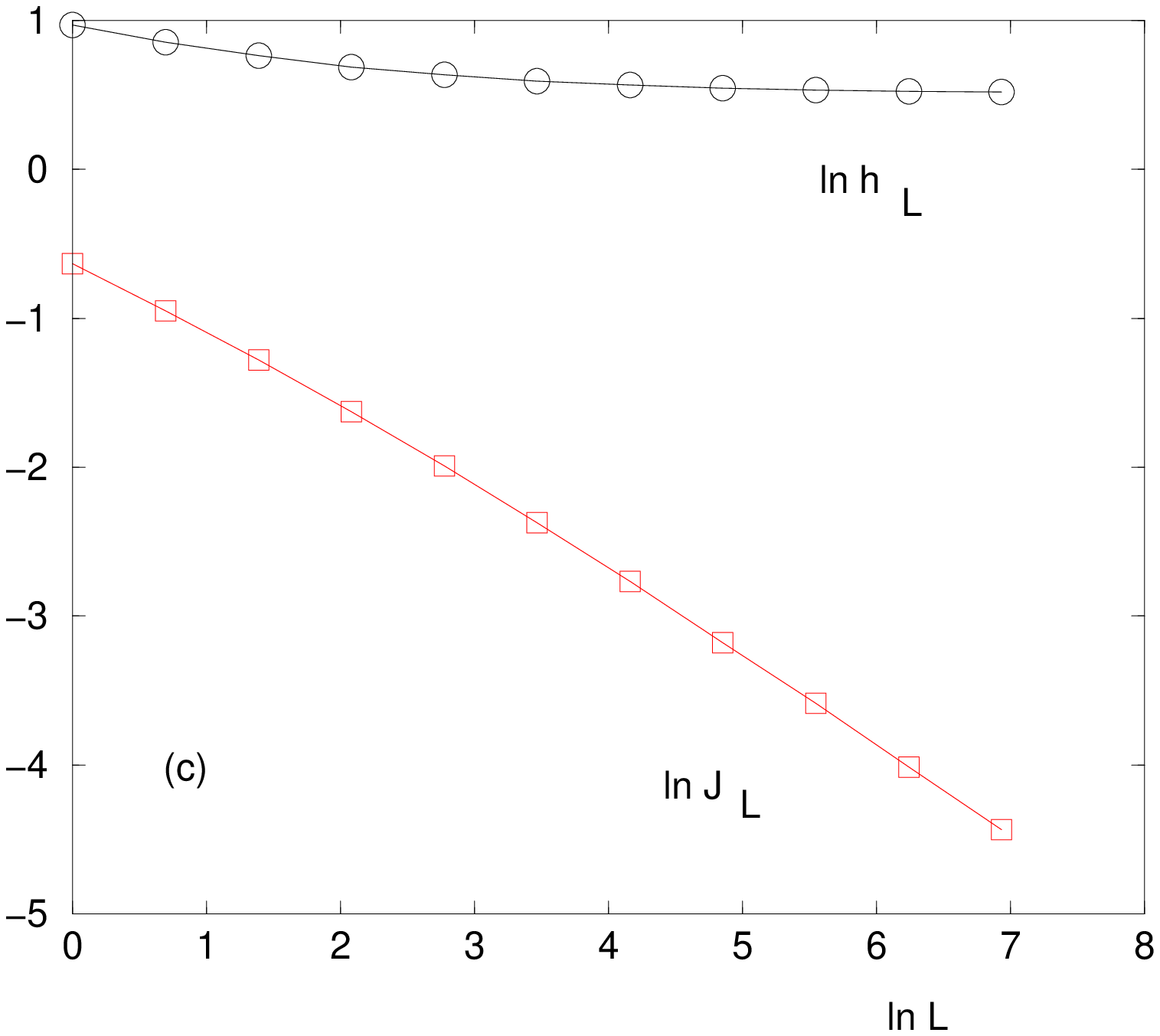}
\caption{ RG flows in log-log scale of the typical renormalized transverse field $ h_L^{typ} $ (circles) and of the renormalized coupling $J_L^{typ} $ (squares) for $\sigma=0.625$ : \\
(a) in the spin-glass phase (example here with $h=1.0$), 
the coupling grows as $J_L^{typ} \propto \Upsilon(h) L^{1-\sigma}$  whereas the transverse field
decays exponentially $ h_L^{typ} \propto e^{- \frac{L}{\xi}} $.  \\
(b) at the critical point $h_c=1.78$, $J_L^{typ} $ and $ h_L^{typ} $ decay with the same power-law
of exponent $z$ (Eq. \ref{deftypcriti})  \\
(c) in the paramagnetic phase (example here with $h=3.0$), the coupling $J_L^{typ} $ decays,
 whereas $ h_L^{typ} $ converges towards a finite value.  }
\label{figflow}
\end{figure}

\subsection{ Numerical details }

We have applied numerically the renormalization rules to $n_s=13.10^3$
disordered samples containing $N=2^{12} =4096$ spins,
corresponding to $12$ generations.
For each renormalization step corresponding to the lengths $L=2^n$
with $n=1,2,..,12$
we have analyzed the statistical properties of the renormalized transverse fields 
$h_L$ and of the renormalized couplings $J_L$. 
In particular, we have measured the RG flows of the corresponding typical values 
\begin{eqnarray}
\ln h_L^{typ} && \equiv \overline{ \ln h_L } 
\nonumber \\
\ln J_L^{typ} && \equiv \overline{ \ln \vert J_L \vert }  
\label{deftyp}
\end{eqnarray}
as a function of the length $L$ for various values 
of the initial transverse field $h$ that represents the control parameter
of the spinglass-paramagnetic transition.

\subsection{ Location of the critical point and measure of the dynamical exponent $z$  }

The critical point $h_c$ corresponds to the unstable fixed point where
the transverse fields and the couplings remain in competition at all scales (see Fig. \ref{figflow}).
When this happens, both typical values of Eq. \ref{deftyp} decay
as a power-law with the dynamical exponent $z$
\begin{eqnarray}
\ln h_L^{typ}(h_c) &&  \propto -z \ln L
\nonumber \\
\ln J_L^{typ}(h_c) &&    \propto -z \ln L
\label{deftypcriti}
\end{eqnarray}
For the four values $\sigma=0.55 $, $\sigma=5/8=0.625$, $\sigma=0.75$ and $\sigma=0.9$ that we have studied,
our numerical data for the critical point $h_c$ and for the dynamical exponent
are given in the Table \ref{table}.

\begin{table}[htbp]
\centerline{
\begin{tabular}{|l|l|l|l|l|l|l|}   \hline
  $\sigma$ & Critical Point $h_c$ & Dynamical exponent $z$    & Stiffness modulus exponent $\mu$& Correlation length exponent $\nu$ & $\omega$ & Gap exponent g \\ \hline
 0.55     & 2.04  &  0.31  & 1.9  & 2.5 & 0.6 &  0.77 \\ \hline
 0.625    & 1.78  &  0.36  & 1.72 & 2.34 & 0.62 & 0.84 \\ \hline
 0.75     & 1.46  &  0.46  & 1.52 & 2.14 & 0.62 & 0.98 \\ \hline
 0.9      & 1.19  &  0.58  & 1.3  & 1.9  & 0.6 & 1.1 \\ \hline
\end{tabular}
}

\caption{Numerical estimates of the critical point $h_c$, of the dynamical exponent $z$, of the stiffness modulus exponent $\mu$, of the correlation length exponent $\nu$, of $\omega$ and of the gap exponent $g$  for four values of the parameter  $\sigma$.}
\label{table}
\end{table}

\subsection{ Finite-size scaling of the renormalized typical coupling }

We now consider the finite-size scaling form of the typical renormalized coupling
\begin{eqnarray}
  J_L^{typ}   && \simeq L^{-z} \Phi \left[ (h-h_c)L^{\frac{1}{\nu}}  \right]
\label{jtypfss}
\end{eqnarray}

\begin{figure}[htbp]
 \includegraphics[height=6cm]{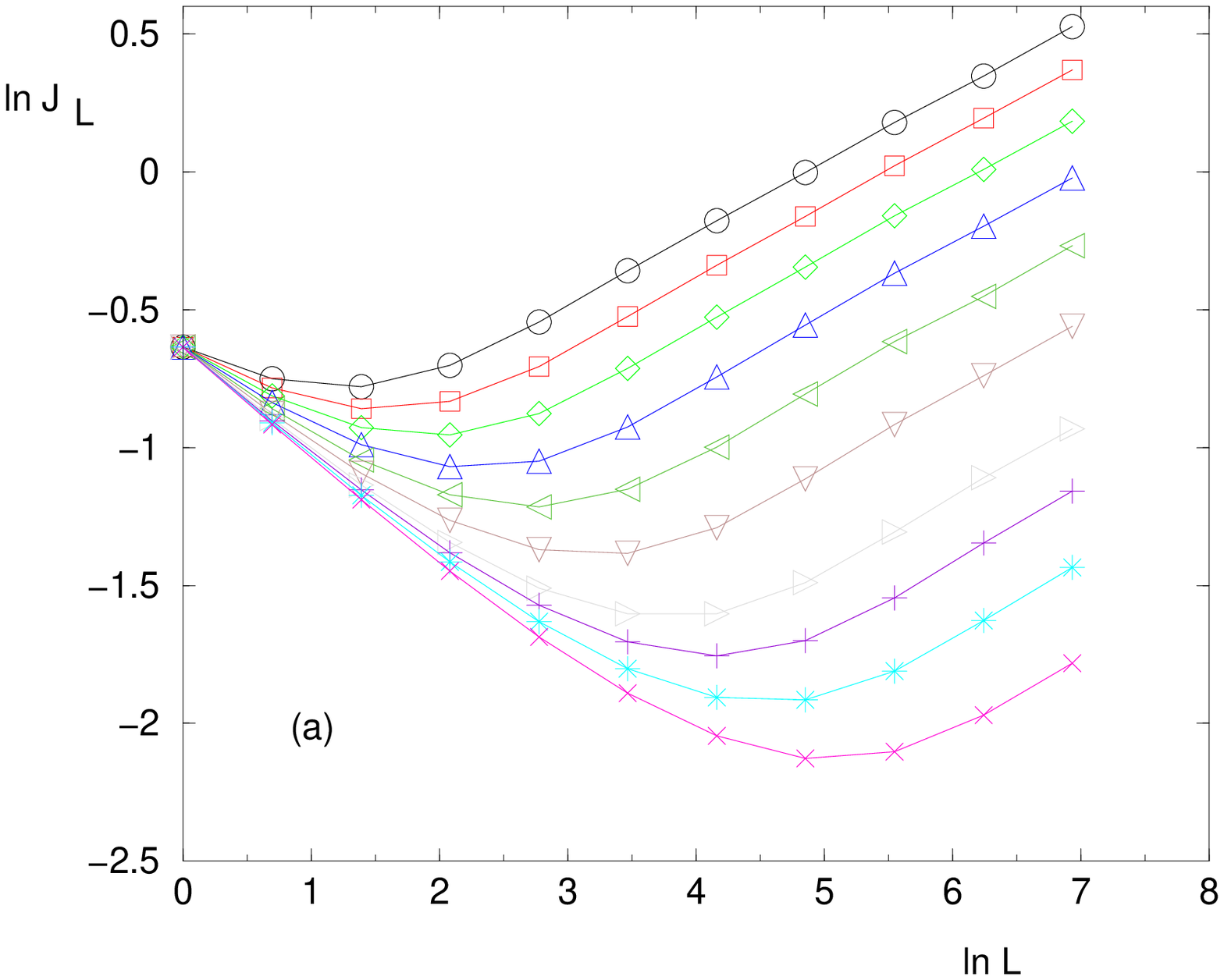}
\hspace{1cm}
 \includegraphics[height=6cm]{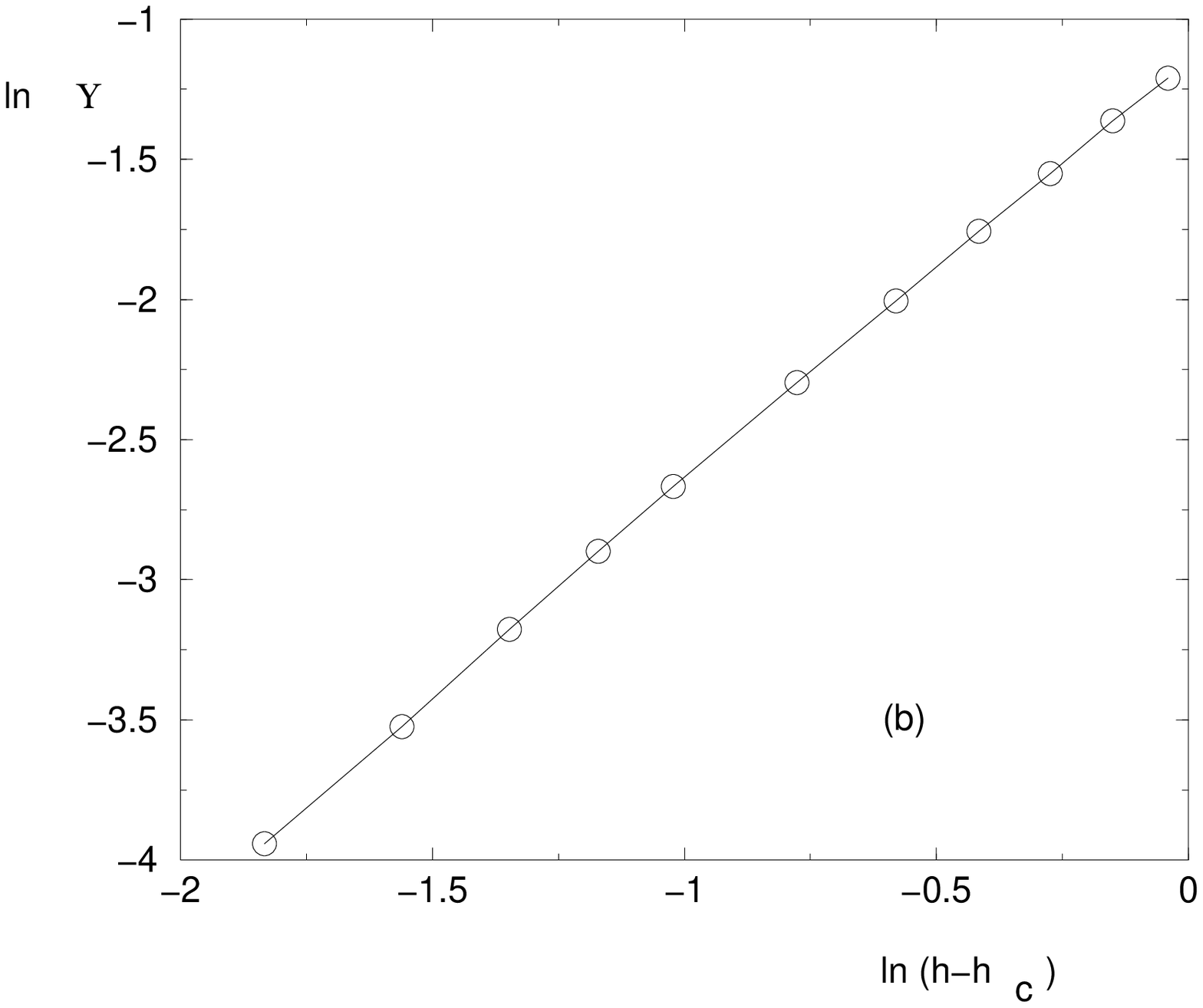}
\caption{ Critical behavior of the stiffness modulus  (Eq. \ref{upsilonh}) for $\sigma=0.75$ : \\
(a) RG flow in log-log scale of the renormalized coupling 
for various values of the initial transverse field $h$ 
in the spin-glass phase $h<h_c$ [namely 
$h=0.5$ (circle);
$h=0.6$ (square);)
$h=0.7$ (diamond);
$h=0.8$ (triangle up);
$h=0.9$ (triangle left);
$h=1.0$ (triangle down);
$h=1.1$ (triangle right);
$h=1.15$ (plus);
$h=1.2$ (star);
$h=1.25$ (cross)]
 :
the asymptotic growth as $\ln J_L^{typ}= (1-\sigma) \ln L + \ln \Upsilon(h) $ (Eq. \ref{jtypsg})
allows to extract the stiffness modulus $\Upsilon(h) $ as a function of the control parameter $h$
 \\
(b) 
$ \ln\Upsilon(h)$ as a function of $\ln (h_c-h)$ yields the slope $\mu \simeq 1.52$ 
 }
\label{figflowjsg}
\end{figure}

\begin{figure}[htbp]
 \includegraphics[height=6cm]{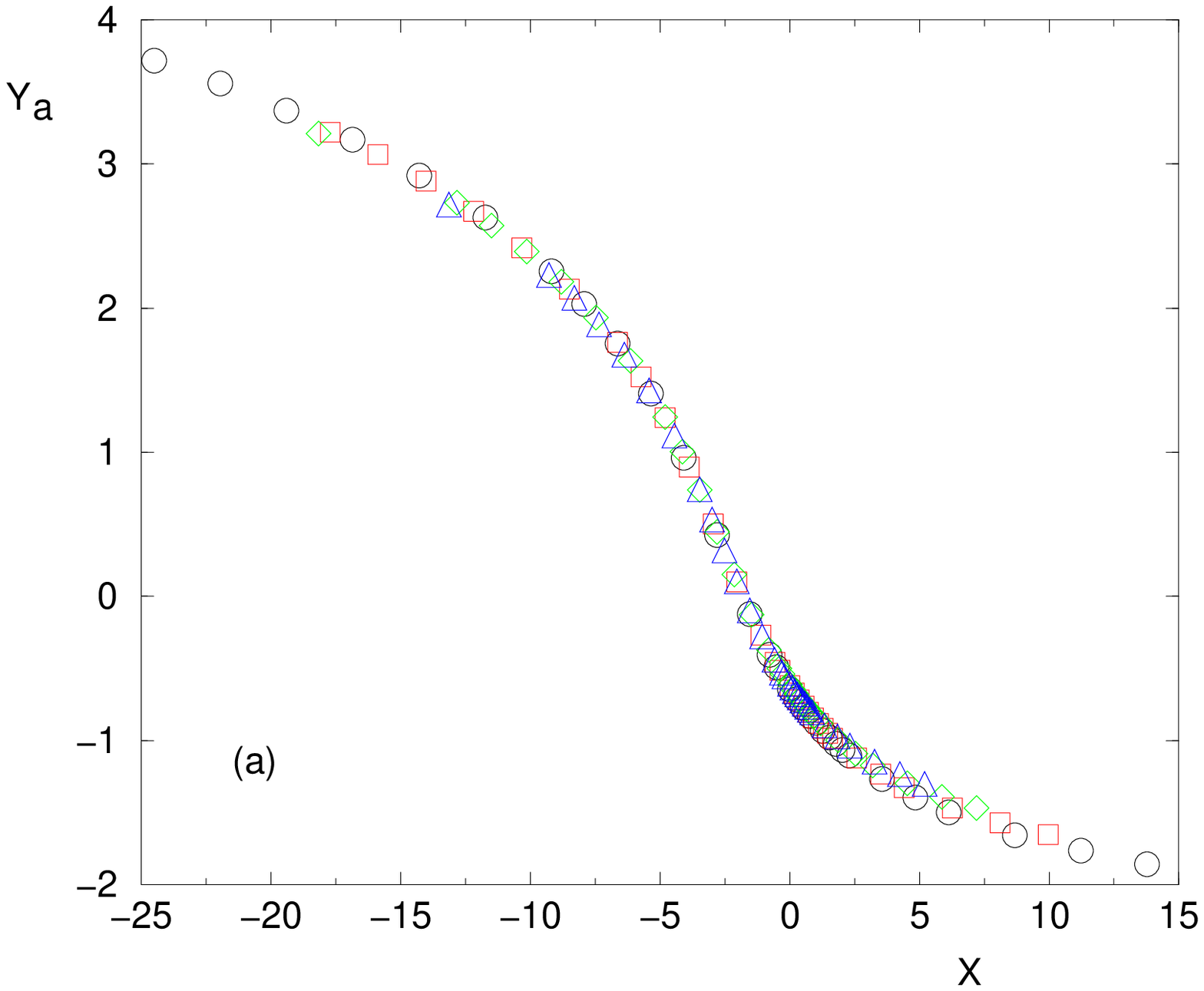}
\hspace{1cm}
 \includegraphics[height=6cm]{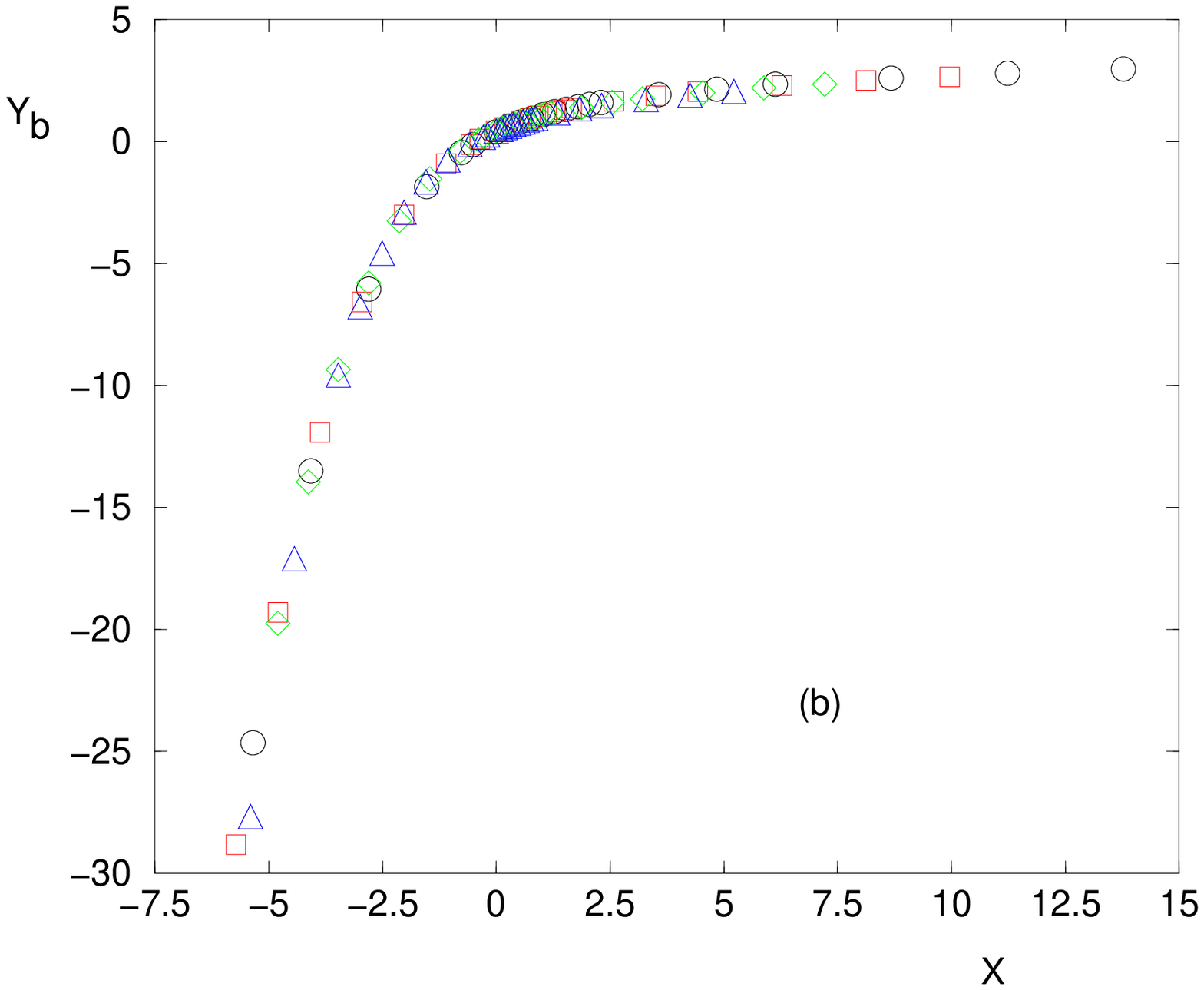}
\caption{ Finite-size scaling for $\sigma=0.75$ with the dynamical exponent $z =0.46 $ and 
the correlation length exponent $\nu=2.14 $ 
shown here for the four bigger sizes $L=4096$ (circles), $L=2048$ (squares), $L=1024$ (diamond)
and $L=512$ (triangles) \\ 
(a) Test of Eq \ref{jtypfss} : $Y_a=\ln J_L^{typ}+z \ln L$ as a function of $X=(h-h_c)L^{\frac{1}{\nu}}  $
 \\
(b)  Test of Eq \ref{htypfss} : $Y_b=\ln h_L^{typ}+z \ln L$ as a function of $X=(h-h_c)L^{\frac{1}{\nu}}  $
 }
\label{figfss}
\end{figure}

In the spin-glass phase $h<h_c$, the typical value of the renormalized coupling
grows as a power-law of exponent $\theta=1-\sigma$ (Eq. \ref{theta})
\begin{eqnarray}
  J_L^{typ} \vert_{h<h_c}  && \oppropto_{L \to +\infty} \Upsilon(h) L^{1-\sigma}
\label{jtypsg}
\end{eqnarray}
where the stiffness modulus $\Upsilon(h) $ vanishes as a power-law
\begin{eqnarray}
\Upsilon(h) && \oppropto_{h \to h_c} (h_c-h)^{\mu} 
\label{upsilonh}
\end{eqnarray}
The matching between Eq. \ref{deftypcriti} and Eq. \ref{jtypsg}
via the finite-size scaling form of Eq. \ref{jtypfss}
yields the relation between critical exponents
\begin{eqnarray}
 \frac{\mu}{\nu} = 1-\sigma+z
\label{relationmu}
\end{eqnarray}
Our numerical data in the spin-glass phase (see Fig. \ref{figflowjsg}) follow the power-laws of
 \ref{jtypsg} and \ref{upsilonh} with the exponent $\mu$ given in Table \ref{table}.
The corresponding numerical values for the correlation length exponent $\nu=\mu/(1-\sigma+z)$ (Eq. \ref{relationmu}) are also given in Table \ref{table}.
We find that these values yield very satisfactory finite-size scaling 
with Eq. \ref{jtypfss} of our numerical data  (see for instance Fig. \ref{figfss} for the case $\sigma=0.75$).

In the paramagnetic phase $h>h_c$, the 
typical renormalized coupling keeps the initial power-law behavior (Eq. \ref{varrgjboxparainte})
\begin{eqnarray}
  J_L^{typ} \vert_{h>h_c}  &&  \oppropto_{L \to +\infty} \frac{A(h)}{L^{\sigma} }
\label{jtypflowpara}
\end{eqnarray}
where the amplitude diverges at criticality
\begin{eqnarray}
 A(h)   &&  \oppropto_{h\to h_c} (h-h_c)^{-\omega}
\label{defomega}
\end{eqnarray}
The matching between \ref{deftypcriti} and Eq. \ref{jtypflowpara}
finite-size scaling form of Eq. \ref{jtypfss}
via the finite-size scaling form of Eq. \ref{jtypfss}
yields the relation
\begin{eqnarray}
\frac{\omega}{\nu}+z =\sigma
\label{relationomega}
\end{eqnarray}
The previous estimates lead to the values $\omega=\nu(\sigma-z)$
(see Table \ref{table}).

\subsection{ Finite-size scaling of the renormalized transverse field  }

The typical renormalized transverse field is expected to
follow the following finite-size scaling form analog to Eq. \ref{jtypfss}
\begin{eqnarray}
  h_L^{typ}   && \simeq L^{-z} \Psi \left[ (h-h_c)L^{\frac{1}{\nu}}  \right]
\label{htypfss}
\end{eqnarray}
The corresponding finite-size scaling of our numerical data 
is shown on Fig. \ref{figfss} (b) for the case $\sigma=0.75$.

In the Spin-Glass phase $h<h_c$, the leading exponential decay of the 
typical transverse field (Eq. \ref{rghstronglogav}) 
allows to define some correlation length $\xi(h)$
\begin{eqnarray}
 \ln h_L^{typ} \vert_{h<h_c}  && = \overline{ \ln h_L } \oppropto_{L \to +\infty} - \frac{L}{\xi(h)} 
\label{htypflow}
\end{eqnarray}
The divergence
near criticality involve the correlation exponent $\nu$ of Eq. \ref{htypfss}
\begin{eqnarray}
\xi(h) && \oppropto_{h \to h_c} (h_c-h)^{- \nu} 
\label{xinuh}
\end{eqnarray}

In the paramagnetic phase $h>h_c$, the 
typical renormalized transverse field converges towards a finite asymptotic value
\begin{eqnarray}
  h_L^{typ} \vert_{h>h_c}  &&  \oppropto_{L \to +\infty} h_{\infty}^{typ}(h) 
\label{htypflowpara}
\end{eqnarray}
that vanishes as a power-law
\begin{eqnarray}
h_{\infty}^{typ} && \oppropto_{h \to h_c} (h-h_c)^{g} 
\label{defg}
\end{eqnarray}
The matching between Eq. \ref{htypfss} and Eq. \ref{htypflowpara}
yields the standard relation for the gap critical exponent
\begin{eqnarray}
g=z \nu
\label{relationg}
\end{eqnarray}
The previous estimates lead to the values given in Table \ref{table}.

\subsection{ Link with the standard critical exponents $(\alpha,\beta,\gamma,\eta)$ }

Up to now, we have described the critical exponents involved in the RG flows of
the transverse fields and of the couplings. It seems now useful to describe the link with the standard critical exponents $(\alpha,\beta,\eta,\gamma)$ 
of the general theory of critical phenomena \cite{thill,rieger} :

(i) the exponent $\alpha$ governing the singular part of the ground state energy
\begin{eqnarray}
e_{GS} \propto \vert h-h_c \vert^{2-\alpha}
\label{egs}
\end{eqnarray}
can be obtained via the quantum hyperscaling relation 
involving the spatial dimensionality $d=1$ and the dynamical exponent $z$
\cite{thill}
\begin{eqnarray}
2 -\alpha= \nu (d+z)= \nu (1+z)
\label{hyperscal}
\end{eqnarray}
Our numerical data leads to the negative values given in Table \ref{tablebis}
for the exponent $\alpha$.

(ii) the exponent $\beta$ governing the vanishing of the Edwards-Anderson order parameter
\begin{eqnarray}
q_{EA} \equiv \overline{ < \sigma_i^x>^2 } \propto \vert h-h_c \vert^{\beta}
\label{qEA}
\end{eqnarray}
yields the following finite-size scaling decay at criticality
\begin{eqnarray}
q_{EA}(L) \propto L^{- \frac{\beta}{\nu} } 
\label{qEAfss}
\end{eqnarray}
The relation with the scaling of the renormalized coupling
\begin{eqnarray}
\overline{J^2(L)} \propto L^{-2z} \simeq J^2_{ini}(L) (L q_{EA} )^2 \propto L^{-2 \sigma +2 -2 \frac{\beta}{\nu}}
\label{qEAfssj}
\end{eqnarray}
yields the relation
\begin{eqnarray}
\frac{\beta}{\nu}=1- \sigma +z
\label{betanu}
\end{eqnarray}
The comparison with Eq. \ref{relationmu} shows that $\beta$ actually coincides
with the exponent $\mu$ of the stiffness modulus
\begin{eqnarray}
\beta = \mu
\label{betamu}
\end{eqnarray}

(iii) the correlation exponent $\eta$ governing the power-law decay of the equal-time spatial correlation function at criticality \cite{thill,rieger}
\begin{eqnarray}
C(r) \equiv \overline{ < \sigma_i^x \sigma_{i+r}^x>^2  } \propto r^{- (d+z-2+\eta)}
\label{corre}
\end{eqnarray}
is related by finite-size scaling to the sqare of
the order parameter $q_{EA}^2(L)\propto L^{- 2\frac{\beta}{\nu} } $ 
leading to
\begin{eqnarray}
d+z-2+\eta = 2 {\beta}{\nu}= 2 (1- \sigma +z) 
\label{correfss}
\end{eqnarray}
so that here with $d=1$, one obtains
\begin{eqnarray}
\eta =  3 - 2\sigma + z 
\label{eta}
\end{eqnarray}

(iv) the experimentally measurable non-linear susceptibility 
  $\chi_{nl}$ scales as 
\begin{eqnarray}
\chi_{nl} \propto L^{\frac{\gamma_{nl}}{\nu}}  
\label{chinl}
\end{eqnarray}
with \cite{rieger,guo}
\begin{eqnarray}
\frac{\gamma_{nl}}{\nu}=2-\eta +2 z = (2 \sigma-1) +z
\label{gammanl}
\end{eqnarray}
whereas the overlap susceptibility involves the exponent \cite{rieger}
\begin{eqnarray}
\frac{\gamma_{overlap}}{\nu}=2-\eta + z = 2 \sigma-1
\label{gammaoverlap}
\end{eqnarray}
and the spin-glass susceptibility involves \cite{rieger,guo}
\begin{eqnarray}
\frac{\gamma_{SG}}{\nu}=2-\eta  = 2 \sigma-1-z
\label{gammaSG}
\end{eqnarray}
We refer to References \cite{rieger,guo} for more details on these various susceptibilities
involving different numbers of integration in the time direction.

(v) the time-autocorrelation decays at criticality as the power-law
\begin{eqnarray}
 \overline{ < \sigma_i^x(0) \sigma_{i}^x(t)>  } \propto t^{- \rho}
\label{corretime}
\end{eqnarray}
with the finite-size-scaling value
\begin{eqnarray}
 \rho = \frac{\beta}{\nu z }= 1 +\frac{1-\sigma}{z}
\label{rho}
\end{eqnarray}

The numerical results of the RG flows given in Table \ref{table} thus translate into the values given in Table \ref{tablebis} for the standard exponents of phase transitions just described.

\begin{table}[htbp]
\centerline{
\begin{tabular}{|l|l|l|l|l|l|l|l|}   \hline
  $\sigma$ &  $\alpha$ &  $\beta$    & $\eta$&  $\gamma_{nl}$ & $\gamma_{overlap}$ & $\gamma_{SG}$ & $\rho$ \\ \hline
 0.55     & -1.27    & 1.9 & 2.21  & 0.41  & 0.1  & -0.21 & 2.45  \\ \hline
 0.625    & -1.18  & 1.72 &  2.11  & 0.61  & 0.25 & -0.11 & 2.04 \\ \hline
 0.75     & -1.12  & 1.52 & 1.96   & 0.96  & 0.5  & 0.04 & 1.54 \\ \hline
 0.9      &  -1.0   & 1.3  & 1.78  & 1.38  & 0.8  & 0.22 & 1.17  \\ \hline
\end{tabular}
}
\caption{Numerical estimates of the standard exponents of the phase transition as obtained via scaling relations from the numerical results of Table \ref{table}.}
\label{tablebis}
\end{table}

\section{ Conclusion }

\label{sec_conclusion}

In this paper, we have studied via real-space renormalization the ground state of the
 Dyson hierarchical version of the quantum Long-Ranged Spin-Glass with
the power-law decaying variance $\overline{J^2(r)} \propto r^{-2\sigma}$.
In particular, we have focused on the spinglass-paramagnetic zero temperature quantum phase transition driven by the initial uniform transverse field $h$.
 In the spinglass phase $h<h_c$, the typical renormalized coupling grows with the length scale $L$ as the power-law $J_L^{typ}(h) \propto \Upsilon(h) L^{\theta}$ with the classical droplet exponent $\theta=1-\sigma$ whereas the typical renormalized transverse field decays exponentially $ h^{typ}_L(h) \propto e^{- \frac{L}{\xi}}$. At the critical point $h=h_c$, the typical renormalized coupling $J_L^{typ}(h_c) $ and the typical renormalized transverse field $ h^{typ}_L(h_c)$ display the same power-law behavior $L^{-z}$ with a finite dynamical exponent $z$. The RG rules have been applied numerically to chains containing $L=2^{12}=4096 $ spins in order to measure the critical exponents for various values of $\sigma$ in the interesting region $1/2<\sigma<1$.

We hope that the present work will motivate other studies on the quantum SpinGlass-Paramagnetic transition at zero temperature in the presence of Long-Ranged couplings. In particular it would be very interesting the compare with values of critical exponents obtained via Monte-Carlo simulations or other numerical methods.

\end{document}